\documentstyle[11pt]{article}
   \newcommand {\nc}{\newcommand}
   \nc{\eq}{\begin{equation}}
   \nc{\en}{\end{equation}}
   \nc{\eqa}{\begin{eqnarray}}
   \nc{\ena}{\end{eqnarray}}
   \nc{\eqann}{\begin{eqnarray*}}
   \nc{\enann}{\end{eqnarray*}}
   \def\prf{{\bf{Proof:}}\\}
   \def\endprf{${\bf{\Box}}$\\}
   \newtheorem{definition}{Definition}
   \newtheorem{lemma}{Lemma}
   \newtheorem{theorem}{Theorem}
   \newtheorem{proposition}{Proposition}
   \newtheorem{corollary}{Corollary}

   \nc {\dfn}[1]{{\it{#1}}}
   \nc{\nn}{\nonumber}
   \def\dlt{\delta}
   \def\ep{\epsilon}
   \def\lam{\lambda}
   \def\th{theory}
   \def\thv{theory }
   \def\gp{group}
   \def\gpv{group }

   \def\subg{sub\gp}
   \def\subgv{sub\gpv }
   
   \def\sbv{sub\gpv }
   
   \def\nsgv{normal \subgv}
   
   \def\stbsbv{stationary \sbv}

   \def\symgv{symmetry \gpv}
   \def\ltgp{little \gp}
   \def\ltgpv{little \gpv}
   
   \def\dgv{double \gpv}
   \def\rep{representation}
   \def\repv{representation }
   \def\Rep{Representation}
   \def\Repv{Representation }
   \def\repth{\repv \th}
   \def\repthv{\repv \thv}
   \def\indr{induced \rep}

   \def\prjrepv{projective \repv}
   \def\svrep{single-valued \rep}
   \def\svrepv{single-valued \repv}
   \def\tvrep{two-valued \rep}
   \def\tvrepv{two-valued \repv}
   \def\spinrep{spinor \rep}
   \def\spinrepv{spinor \repv}
   \def\md{module}
   \def\mdv{module }
   \def\irr{irreducible}
   \def\irrv{irreducible }
   \def\id{equivalent}
   \def\idv{equivalent }

   \def\conj{conjugate}
   \def\conv{conjugate }
   
   \def\smdpv{semi-direct product }
   \def\dim{dimension}
   \def\diml{\dim al}
   \def\dimv{dimension }
   \def\dimlv{\dim al }
   
   \def\fdimlv{four-\dimlv}
   \def\fdim{four \dim}
   \def\fdimv{four \dimv}

   \def\strv{structure }
   \def\hep{high energy physics}

   \def\wpv{wreath product }
   \def\etal{{\it et al} }

   \def\decpv{decomposition }
   
   \def\clsfv{classification }

   \nc{\sqt}{\sqrt{2}}
   \nc{\tsqt}{2\sqt}
   \nc{\msqt}{$\sqt$}
   \nc{\nsqt}{$-\sqt$}
   \nc{\mtsqt}{$\tsqt$}
   \nc{\ntsqt}{$-\tsqt$}

   \def\ot{\otimes}
   
   \def\smdp{>\hspace{-0.2cm}\lhd}
   \def\otl{\bigotimes\limits}
   \def\prodl{\prod\limits}
   \def\suml{\sum\limits}
   \nc {\inv}[1]{#1^{-1}}
   \nc {\hc}[1]{{#1}^\dag}
   \nc {\cc}[1]{{#1}^\ast}
   \nc {\ad}[2]{Ad_{#1}({#2})}
   \nc {\wad}{\widetilde{Ad}}
   \nc {\wadf}[2]{\widetilde{Ad}_{#1}({#2})}
   \nc {\pb}[1]{{#1}^\ast}
   \def\tld{\tilde}
   \def\pr{\prime}
   \def\ov{\overline}
   \def\indxst{{\cal I}}
   \def\intg{{\cal Z}}
   \def\real{{\cal R}}
   \def\complex{{\cal C}}
   \def\quaternion{{\bf{H}}}
   \nc {\ga}[2]{{#1}[{#2}]}
   \nc {\cga}[1]{\ga{\complex}{#1}}
   \nc {\cgag}{\cga{G}}
   \nc {\eu}[1]{E^{#1}}
   \nc {\euf}{\eu{4}}
   \nc {\eun}{\eu{n}}
   \nc {\zn}[1]{\intg^{#1}}
   \nc {\zf}{\zn{4}}
   \nc {\zt} {Z_2}
   \nc {\ztn}[1]{\zt^{#1}}
   \nc {\ztt} {\ztn{2}}
   \nc {\ztth} {\ztn{3}}
   \nc {\ztf} {\ztn{4}}
   \nc {\ztnf}{\ztn{n}}
   \nc {\per}[1]{S_{#1}}
   \nc {\pern}{\per{n}}
   \nc {\irrr}[2]{IRR_{#1}({#2})}
   \nc {\irrc}[1]{\irrr{\complex}{#1}}
   \nc {\irrcg}{\irrc{G}}
   \def\hf{{1\over 2}}
   \def\ebr{\bar{e}}

   \nc {\mn}{(-)}
   \nc {\mns}[1]{\mn^{#1}}
   \nc {\mo}{(-1)}
   \nc {\mos}[1]{\mo^{#1}}
   \nc {\unit} {{\bf 1}}
   \nc {\unt} {\unit_{2\times 2}}
   \nc {\unth} {\unit_{3\times 3}}
   \nc {\unf} {\unit_{4\times 4}}
   \nc {\une} {\unit_{8\times 8}}
   \nc {\unw} {\unit_{12\times 12}}
   \nc {\zrt} {{\bf 0}_{2\times 2}}
   \nc {\zrth} {{\bf 0}_{3\times 3}}
   \nc {\zrf} {{\bf 0}_{4\times 4}}
    \def\CDalign#1{\bgroup\vcenter\bgroup\tabskip 2pt 
      \baselineskip 14pt \lineskip 3pt \lineskiplimit 3pt
      \halign\bgroup &\hfill$##$\hfill\crcr
      #1\crcr\egroup\egroup\egroup} 
    \def\CDdown{\Big\downarrow}       
    \def\CDrlabel#1{\vcenter{\hbox to0pt{$\scriptstyle#1$\hss}}} 
    \def\CDto{\mathop{\relbar\joinrel\longrightarrow}\limits}    
    \def\CDup{\Big\uparrow}  
    \def\CDeq{\Big\|}
  \nc{\act}[3]{S_{{#1}}^{{#2}}[{#3}]}
  \nc{\cact}[2]{S_{Cl}^{{#1}}[{#2}]}
  
   \nc {\prj}[4]{\pi_{#1,#2}#2\ot_{S_#1}e^o_{#3,#4}}
   \nc {\prjf}[2]{\prj{o}{#1}{\eta}{#2}}
   \nc {\prjff}{\prjf{h}{i}}
   \nc {\Prj}[4]{\Pi_{#1,#2;#3,#4}}
   \nc {\Prjf}[2]{\Prj{o}{#1}{\eta}{#2}}
   \nc {\Prjff}{\Prjf{h}{i}}
   \nc {\orbt}[2]{\pi_{#1,#2}}
   \nc {\orbtf}[1]{\orbt{o}{#1}}
   \nc {\orbte}{\orbtf{e}}
   \nc {\rp}[3]{\orbt{#1}{#2}(#3)}
   \nc {\rpf}[2]{\rp{o}{#1}{#2}}
   \nc {\stbrp}[2]{D^o_#1(\tld{s}(#2))}
   \nc {\stbrpme}[4]{\stbrp{#1}{#2}^{#3}_{#4}}
   \nc {\stbrpf}[1]{\stbrp{\eta}{#1}}
   \nc {\stbrpmef}[3]{\stbrpme{\eta}{#1}{#2}{#3}}
   \nc {\ch}[2]{\chi_{#1;#2}}
   \nc {\che}[3]{\ch{#1}{#2}(#3)}
   \nc {\chf}[1]{\ch{o}{#1}}
   \nc {\chff}{\chf{\eta}}
   \nc {\chef}[2]{\che{o}{#1}{#2}}
   \nc {\cheff}[1]{\chef{\eta}{#1}}
   \nc {\chsb}[1]{\chi^o_{#1}}
   \nc {\chsbe}[2]{\chsb{#1}(#2)}
   \nc {\chsbf}{\chsb{\eta}}
   \nc {\chsbef}[1]{\chsbe{\eta}{#1}}
   \nc {\cg}[1]{O_{#1}}
   \nc {\cgn} {\cg{n}}
   \nc {\oh} {\cg{4}}
   \nc {\ohd} {{\overline{\oh}}}
   \nc {\ztfb} {\overline{\ztf}}
   \nc {\iso}[2]{ISO_{#1}(#2)}
   \nc {\isod}[1]{\iso{d}{#1}}
   \nc {\eg}[1]{ISO(#1)}
   \nc {\egn}{\eg{n}}
   \nc {\fs}[1]{F(#1)}
   \nc {\fsf}{\fs{\emb}}
   \nc {\fx}[1]{I(#1)}
   \nc {\fxf}{\fx{\emb}}
   \nc {\wrn}[1]{\ztn{#1}\smdp\per{#1}}
   \nc {\wn}{\wrn{n}}
   \nc {\wnn}{\pern^{\zt}}
   \nc {\w}[1]{\per{#1}^{\zt}}
   \nc {\fix}[1]{\per{(n-#1)}\ot \per{p}}
   \nc {\fixp}{\fix{p}}
   \nc {\cb}[1]{C_{#1}}
   \nc {\cn}{\cb{n}}

   \def\hpcbcv{hyper-cubic }
   
   \def\cbgv{cubic \gpv}
   \nc {\prm}{\sigma}
   \nc {\prmt}{\tld{\prm}}
   \nc {\prmp}{\prm^\pr}
   \nc {\cyc}[2]{\tau_{{#1}{#2}}}
   \nc {\emb}{\iota}
   \nc {\epy}{\tld{\ep}}
   \nc {\de}[1]{d_{E^{#1}}}
   \nc {\den}{\de{n}}
   \nc {\dy}{\tld{d}}
   \nc {\repw}[2]{e_{(#1)#2}}
   \nc {\repww}[4]{\repw{#1}{#2}\ot\repw{#3}{#4}}
   \nc {\prw}[6]{\pi_{#1,#2}#2\ot_{F_#1}(\repww{#3}{#4}{#5}{#6})}
   \nc {\dm}[1]{d_{({#1})}}
   \nc {\sls}{{\it slash}}
   \nc {\slsv}{{\it slash} }
   \nc {\lcl} {{\it local}}
   \nc {\lclv} {{\it local} }
   \nc {\drln}[5]
    {\put(#1,#2){\line(#3,#4){#5}}}
   \nc {\ybxa}[4]
    {
     \begin{picture}(40,10)
      \drln {0}{0}{0}{1}{10}
      \drln {0}{0}{1}{0}{40}
      \drln {10}{0}{0}{1}{10}
      \drln {0}{10}{1}{0}{40}
      \drln {20}{0}{0}{1}{10}
      \drln {30}{0}{0}{1}{10}
      \drln {40}{0}{0}{1}{10}
      \drln {0}{0}{1}{1}{#1}
      \drln {10}{0}{1}{1}{#2}
      \drln {20}{0}{1}{1}{#3}
      \drln {30}{0}{1}{1}{#4}
     \end{picture}
    }
   \nc {\ybxb}[4]
    {
     \begin{picture}(30,20)
      \put(0,0){\line(0,1){20}}
      \put(0,0){\line(1,0){10}}
      \put(10,0){\line(0,1){20}}
      \put(0,10){\line(1,0){30}}
      \put(0,20){\line(1,0){30}}
      \put(20,10){\line(0,1){10}}
      \put(30,10){\line(0,1){10}}
      \drln {0}{0}{1}{1}{#1}
      \drln {0}{10}{1}{1}{#2}
      \drln {10}{10}{1}{1}{#3}
      \drln {20}{10}{1}{1}{#4}
     \end{picture}
    }
   \nc {\ybxc}[4]
    {
     \begin{picture}(20,20)
      \put(0,0){\line(0,1){20}}
      \put(0,0){\line(1,0){20}}
      \put(10,0){\line(0,1){20}}
      \put(0,10){\line(1,0){20}}
      \put(0,20){\line(1,0){20}}
      \put(20,0){\line(0,1){20}}
      \drln {0}{0}{1}{1}{#1}
      \drln {10}{0}{1}{1}{#2}
      \drln {0}{10}{1}{1}{#3}
      \drln {10}{10}{1}{1}{#4}
     \end{picture}
    }
   \nc {\ybxd}[4]
    {
     \begin{picture}(20,30)
      \drln {0}{0}{0}{1}{30}
      \drln {0}{0}{1}{0}{10}
      \drln {10}{0}{0}{1}{30}
      \drln {0}{10}{1}{0}{10}
      \drln {0}{20}{1}{0}{20}
      \drln {0}{30}{1}{0}{20}
      \drln {20}{20}{0}{1}{10}
      \drln {0}{0}{1}{1}{#1}
      \drln {0}{10}{1}{1}{#2}
      \drln {0}{20}{1}{1}{#3}
      \drln {10}{20}{1}{1}{#4}
     \end{picture}
    }
   \nc {\ybxe}[4]
    {
     \begin{picture}(10,40)
      \drln {0}{0}{0}{1}{40}
      \drln {10}{0}{0}{1}{40}
      \drln {0}{0}{1}{0}{10}
      \drln {0}{10}{1}{0}{10}
      \drln {0}{20}{1}{0}{10}
      \drln {0}{30}{1}{0}{10}
      \drln {0}{40}{1}{0}{10}
      \drln {0}{0}{1}{1}{#1}
      \drln {0}{10}{1}{1}{#2}
      \drln {0}{20}{1}{1}{#3}
      \drln {0}{30}{1}{1}{#4}
     \end{picture}
    }
   \def\orb{{\bf{Orbit }}}
   \nc{\op}{orientation-preserved}
   \nc{\opv}{orientation-preserved }
   \nc{\on}[1]{SO_{#1}}
   \nc{\ohn}[1]{O_{#1}}
   \nc{\of}{\on{4}}
   \nc{\ohf}{\ohn{4}}
   \nc{\ofd}{\overline{\of}}
   \nc{\ohfd}{\overline{\ohf}}
   \nc{\cir}[1]{e^{i({#1})\pi}}
 \title{Structure and \repv theory of double group of four-dimensional cubic group
 }
 \author{Jian Dai\thanks{daij1492@yahoo.com},
  Xing-Chang Song\thanks{songxc@ibm320h.phy.pku.edu.cn}\\
  Department of Physics, Peking University}
 \date{February 13, 1999\\
   Revised on September 4, 2000}
 
\begin{document}
  \begin{titlepage}
  \maketitle
  \begin{abstract}
  \noindent
   We generalize the concept of cubic group into any \dimv and derive their \conv classifications and \repth s.
   Double \gpv and \spinrepv are defined.
   A detailed calculation is carried out on the structures of four-\dimlv cubic group
   $\oh$ and its double group, as well as all in\idv \svrep s and \spinrep s of $\oh$.
   All \rep s are derived adopting Clifford theory of
   \decpv of \indr s.
  \end{abstract}
  \end{titlepage}
  \section{Introduction}
   It is well-known that electrons stay
   in \spinrep s of \symgv of a given lattice in condensed matter physics;
   it is reasonable to assume that quarks, leptons, as well as baryons, should
   reside in \spinrep s of \symgv of a \fdimlv lattice in
   lattice field theory (we will make the concept
   "\spinrep" precise in the next section). Accordingly, to explore the
   \strv and \rep s (\spinrep s especially) of such groups has important significance in \hep.\\
   \\
   In the following we concentrate on \hpcbcv case, though it
   is not the maximum symmetric lattice in \fdimv \cite{bc}. The first \rep-theoretical
   consideration of \symgv of such lattice was given by A.
   Young \cite{young}. Then mathematicians worked in this field
   due to the interest of \wpv \cite{sp1}\cite{sp2} to which A.
   Kerber gave a thorough review in his book \cite{kerber}.
   Physicists took part in after K. G. Wilson introduced
   lattice gauge theory \cite{wilson}. M. Baake \etal first gave an explicit description of characters
   of \fdimlv \cbgv \cite{baake}; J. E. Mandula \etal derived the same results using a different method
   \cite{mandula}. However, the problem of \spinrep s is still a vacancy.\\
   \\
   In this paper, we make full use of the power of Clifford theory on \decpv of \indr s (Sec.\ref{indrep}).
   We give a systematic and schematic description of \conv \clsfv and \repthv of generalized
   \cbgv $\cgn$ (Sec.\ref{ohstr}). The concept \dgv is introduced in Sec.\ref{doublegroup}
   to clarify the terms "\svrep" and "\tvrep (\spinrep)". Then we
   specify our general results to \fdim, give a detailed description of \strv and \conv \clsfv of $\oh$ and
   its double $\ohd$ (Sec.\ref{strandcc}), and derive all in\idv
   \svrep s as well as \spinrep s of $\oh$, adopting Clifford theory (Sec.\ref{repofohd}).
   We would like point out that the "spinor" part of our work is
   completely new and that although the "single-valued" part is well-known,
   our method to derive them is much more tidy and systematic than
   that used by other authors who gave the same results, thanks for the power of
   Clifford theory.
  \section{Conceptual foundations}
   \subsection{Clifford theory on decomposition of \indr s}
   \label{indrep}
    We will apply two results of Clifford theory, a powerful method in decomposing
    \indr s of a given \gpv $E$ with a \nsgv $N$
    \cite{cr}\cite{s}\cite{cs}\cite{qiu}. We will use $\cga{E}$ for the group algebra of $E$ in complex
    field and $G$ for $E/N$ below. The first result is
    \begin{theorem}(Clifford)\cite{clif}\cite{cr}
    \label{clifford}
     Let $M$ be a simple $\cga{E}$-\md, and $L$ a simple
     $\cga{N}$-sub\mdv of $M_N$ s.t. $L$ is stable relative to
     $E$, i.e. $L$ is isomorphic to all of its \conj s. Then
     \[
      M\cong L\ot_{\complex}I
     \]
     for a left ideal $I$ in $End_{\cga{E}}L^E$. The $E$-action
     on $L\ot_{\complex}I$ is given by
     \[
      x\mapsto U(x)\ot V(x), x\in E
     \]
     where $U:E\rightarrow GL(L)$ is a \prjrepv of $E$ on $L$, and
     $V:E\rightarrow GL(I)$ is a \prjrepv of $G$, that is, $V(x)$
     depends only on the coset $xN$ of $x$ in $G$, for each $x\in
     E$. The factor sets associated with $U$ and $V$ are inverse
     of each other.
    \end{theorem}
    The second result can be viewed as a special case of Theorem
    \ref{clifford}. Let $E=N\smdp G, |E|<\infty$ and $N$ be abelian, then adjoint action
    of $G$ upon $N$ makes $N$ a $G$-module. This $G$-action can be extended naturally
    to a $G$-action upon $\cga{N}$ by linearity.
    Define $\Pi(N):=\{\pi_\mu\}\subset \cga{N}$,
    \eq
    \label{defpro}
     \pi_\mu:=\sum_{a\in N}{\chi_\mu (\inv{a}) a}
    \en
    where $\chi_\mu$ are all \irrv \rep s of
    $N$. The $G$-action on $\Pi(N)$ is closed and thus $\Pi(N)$
    is separated into orbits
    $\Pi(N)=\coprod\limits_{o\in \indxst}{\Pi_o}$ where $\indxst$ is a index set to label
    different orbits. For each $\Pi_o$,
    choose one of its element and denote it as $\orbte$. The
    stablizer of each $\orbte$ in $G$ (\dfn{\ltgp}) is denoted as $S_o$.
    There is a bijection from $G/S_o=\{hS_o \}$ to $\Pi_o$ defined
    by
    \eq
    \label{orbit}
     Ad_h(\orbte)=h\orbte \inv{h}=:\orbtf{h}
    \en
    where $\{h\}$ is a system of representatives of left cosets
    $G/S_o$. Define
    \eq
    \label{main}
     \Prjff\equiv\prjff
    \en
    in which $\{e^o_{\eta,i}|i=1,2,...,d^o_\eta\}$ with fixed $o,\eta$ is the $\eta$th \irrv \repv of
    $S_o$ whose \dimv is $d^o_\eta$, then
    \begin{proposition}(\ltgpv method)\cite{s}\cite{cs}\cite{qiu}
    \label{rfm}
     \begin{enumerate}
      \item For each fixed $(o,\eta)$, $\{\Prjff\}$ induces an \irrv \repv of
       $E$, denoted as $D_{o,\eta}$;
      \item If $(o,\eta)\neq (o^\pr,\eta^\pr)$, then
       $D_{o,\eta}$ and $D_{o^\pr,\eta^\pr}$ are
       in\id;
      \item $\{D_{o,\eta}\}$ gives all in\idv \irrv \rep s of $E$.
     \end{enumerate}
    \end{proposition}
   \subsection{Cubic group in any dimension}
   \label{ohstr}
    The symmetry group of a cube including inversions in three dimensional Euclidean space,
    which is denoted as $O_h$ in the theory of point groups \cite{hs},
    can be generalized into any $n$-dimensional Euclidean space
    $\eun$, along two different approaches whose results are \id.\\
    \\
    The first approach of generalization which is very natural and straightforward
    is geometrical. An \dfn{$n$-cube} (or \dfn{hyper-cube in $\eun$}) $\cn$ is defined to
    be a subset of $\eun$, $\cn=\{p|x^i(p)=\pm 1\}$, where $x^i:\eun\rightarrow \real, i=1,2,...,n$
    are coordinate functions of $\eun$, together with the distance inherited from $\eun$.
    \dfn{$n$-Cubic group} (\dfn{hyper-cubic group of degree $n$}) $O_n$ consists of all isometries of $\eun$ which stabilize
    $\cn$. While the second approach of generalization is
    algebraic. $O_h$ has a \smdpv structure
    as $\ztth\smdp\per{3}$ \cite{s}; we generalized this to
    $\wn$ which is just a \dfn{wreath product} $\zt\wr\pern$ of
    $\zt$ with $\pern$. We point out that these two
    generalizations are identical. Let $\{e_i\}$ be
    a standard orthogonal basis of $\eun$, namely $x^j(e_i)=\dlt^j_i$.
    Define $n$ points in $\cn$ to be $p_0=(-1,-1,...,-1),
    p_i=p_0+2e_i$.
    \begin{lemma}
     $\forall \ep \in \cgn$, $\ep$ is entirely determined by
     images $\ep(p_i), i=0,1,2,...,n$.
    \end{lemma}
    \prf
     The fact that $\ep$ is an isometry of $\eun$ ensures the equality of Euclidean distances $d(p,p_i)=d(\ep(p),\ep(p_i)), i=0,1,...,n$
     for any other $p$ in $\cn$.
     If all $\ep(p_i)$ are given, $\ep(p)$ will be fixed for any other $p$
     accordingly due to the fundamental lemma of Euclidean geometry (lemma
     \ref{fundament} in Appendix \ref{app2}). In fact, the existence of solution in lemma \ref{fundament}
     is guaranteed by that $\ep$ stablizes $\cn$ and
     lemma \ref{fundament} itself ensures the uniqueness.\\
    \endprf
    To fix $\ep(p_0)$, there are $2^n$ ways; while for a fixed $\ep(p_0)$, there are $n!$
    possibilities to fix $\ep(p_i), i=1,2,...,n$. Therefore, $|\cgn|=2^n\cdot n!$.
    \begin{proposition}(Structure of $\cgn$)
    \label{On}
     \eq
     \label{Oneq}
      \cgn\cong \zt^n\smdp S_n
     \en
    \end{proposition}
    \prf
     Introduce a class of isometries in $\eun$:
     \eq
     \label{cong}
      \prm(e_i)=e_{\prm(i)}; I_i(e_j)=(1-2\dlt_{ij})e_j,
      i=1,2,...,n
     \en
     where $\prm\in \pern$ permute the axes and $I_i$ inverts the $i$th
     axis. Subjected to the relations
     \eq
     \label{genr}
      I_i^2=e, I_iI_j=I_jI_i, i,j=1,2,...,n;
      \prm I_i=I_{\prm(i)}\prm, \prm\in\pern
     \en
     these isometries generate a sub-group of $\cn$ isomorphic to $\wn$ whose order
     is $2^n n!=(|\cgn|)$. So (\ref{Oneq}) follows.\\
    \endprf
    A. Kerber gave a detailed introduction on the \conv classification and \repthv
    of a general \wpv $N\wr G$ in \cite{kerber}. We specify his general results to our case $\wn\cong \zt\wr \pern$.\\
    \\
    We recall some fundamental facts about symmetrical \gp s $\pern$ \cite{hs}.
    Each element $\prm\in \pern$ has a cycle decomposition
    \eq
    \label{prmdc}
     \prm=
     \left(
       \begin{array}{cccc}
        1      &2      &\ldots&n      \\
        \prm(1)&\prm(2)&\ldots&\prm(n)
       \end{array}
      \right)
      =\prodl_{k=1}^n\prodl_{\alpha=1}^{\nu_k}\cyc{k}{\alpha}
    \en
    where $\cyc{k}{\alpha}$ are independent $k$-cycles, which can be expressed as $(a_1a_2...a_k)$, and write $n(k,\alpha)
    =\{a_1,a_2,...,a_n\}$. The cycle structure of $\prm$ can be
    represented formally as
    \eq
    \label{sh}
     (\nu)=\prodl_{k=1}^n{(k^{\nu_{k}})}
    \en
    where $\{\nu_k\}$ satisfies $\suml_{k=1}^n{k\cdot \nu_k}=n$.
    Two elements in $\pern$ are \conv \id, iff they have the same cycle
    structure. The number of elements in class
    $(\nu)$ is equal to $N_{(\nu)}=n!/\prodl_{k=1}^n{(k^{\nu_k}\nu_k!)}$.
    Each cycle structure $(\nu)$ can be visualized by one unique Young diagram which is
    denoted also by $(\nu)$.
    There is a one-one correspondence between all in\idv \irrv
    \rep s of $\pern$ and all Young diagrams, which enable us to represent each \irrv \repv
    by the corresponding Young diagram $(\nu)$. We write the basis of one of these \rep s $(\nu)$ in $\dm{\nu}$ \dimv
    as $\repw{\nu}{i}, i=1,2,...,\dm{\nu}$.
   \\
   \\
    We point out that the \conv \clsfv of $\cgn$ has a deep relation with that of
    $\pern$. A generic element in $\zt\wr\pern$ can be written as
    \eq
    \label{cycle}
     \prm \cdot\prodl_i{I_i^{s_i}}=
     \left(
       \begin{array}{cccc}
        1                 &2                 &\ldots&n                \\
        \mns{s_1}\prm(1)&\mns{s_2}\prm(2)&\ldots&\mns{s_n}\prm(n)
       \end{array}
     \right)
    \en
    in which $s_i\in\intg/2\intg$. We call the r.h.s. of (\ref{cycle}) by
    {\it permutation with signature}.
    $\prm\prodl_i{I_i^{s_i}}$ can be
    decomposed according to (\ref{prmdc}), i.e.$\prodl_i{I_i^{s_i}}=\prodl_{k=1}^n\prodl_{\alpha=1}^{\nu_i}\prodl_{a\in n(k,\alpha)}
    I_a^{s_a}$ and
    \eq
    \label{decompwnn}
     \prm\prodl_i{I_i^{s_i}}=\prodl_{k=1}^n\prodl_{\alpha=1}^{\nu_i}(\cyc{k}{\alpha}\prodl_{a\in n(k,\alpha)}
     I_a^{s_a})
    \en
    The {\it cycle with signature} is defined to be
    \[
    \label{cycle1}
     \cyc{k}{\alpha}\prodl_{a\in n(k,\alpha)}I_a^{s_a}=
     \left(
       \begin{array}{cccc}
        a_1              &a_2              &\ldots &a_k            \\
        \mns{s_{a_1}}a_2 &\mns{s_{a_2}}a_3 &\ldots &\mns{s_{a_k}}a_1
       \end{array}
      \right)
    \]
    For two independent $(k,\alpha),(k^\pr,\alpha^\pr)$, it is easy
    to verify that
    \[
     \cyc{k}{\alpha}\cyc{k^\pr}{\alpha^\pr}=\cyc{k^\pr}{\alpha^\pr}\cyc{k}{\alpha},
     \prodl_{a\in n(k,\alpha)}I_a^{s_a}\cyc{k^\pr}{\alpha^\pr}=\cyc{k^\pr}{\alpha^\pr}\prodl_{a\in
     n(k,\alpha)}I_a^{s_a},
     \prodl_{a\in n(k^\pr,\alpha^\pr)}I_a^{s_a}\cyc{k}{\alpha}=\cyc{k}{\alpha}\prodl_{a\in n(k^\pr,\alpha^\pr)}I_a^{s_a}
    \]
    \begin{proposition}
    \label{ccpro}
    \cite{sp1}\cite{sp2}\cite{kerber}We use $\sim$ to denote \conv
    \id.
     \begin{enumerate}
      \item (descent rule)
       \eq
       \label{des}
        \prm\prodl_i{I_i^{s_i}}\sim \prm^\pr\prodl_i{I_i^{s_i^\pr}}
        \Rightarrow\prm\sim\prm^\pr
       \en
      \item (permutation rule) Let
       \[
        \prmt=
         \left(
          \begin{array}{cccc}
           1       &2       &\ldots&n       \\
           \prmt(1)&\prmt(2)&\ldots&\prmt(n)
          \end{array}
         \right)
         =
         \left(
          \begin{array}{cccc}
           \prm (1)&\prm (2)&\ldots&\prm (n)\\
           \prmp(1)&\prmp(2)&\ldots&\prmp(n)
          \end{array}
         \right)
       \]
       then
       \eq
       \label{cyc}
        \prmt(\prm\prodl_i{I_i^{s_i}})\inv{\prmt}=
         \left(
          \begin{array}{cccc}
           \prmt(1)         &\prmt(2)         &\ldots&\prmt(n)        \\
           \mns{s_1}\prmp(1)&\mns{s_2}\prmp(2)&\ldots&\mns{s_n}\prmp(n)
          \end{array}
         \right)
       \en
      \item (signature rule within one cycle) Let $\cyc{k}{\alpha}$ be a $k$-cycle
       and $a_0$ be a given number in $n(k,\alpha)$, then
       \eq
       \label{signat}
        \cyc{k}{\alpha}\prodl_{a\in n(k,\alpha)}I_a^{s_a}\sim
        \cyc{k}{\alpha}\prodl_{a\in
        n(k,\alpha)}I_a^{s_a+\dlt_{aa_0}+\dlt_{a,\cyc{k}{\alpha}(a_0)}}
       \en
       Note that $\cyc{k}{\alpha}(a_0)$ is calculated modulo $k$ (the subscripts of $I_a$ are always understood in this way).
      \item (signature rule between two cycles) Let $\cyc{k}{\alpha},\cyc{k}{\beta}$
       be two independent $k$-cycles and we define a bijection $\theta:
       n(k,\alpha)\rightarrow n(k,\beta), a_i\mapsto b_i$. Then
       \eq
       \label{exchg}
        \cyc{k}{\alpha}\prodl_{a\in n(k,\alpha)}I_a^{s_a}\cdot\cyc{k}{\beta}\prodl_{b\in n(k,\beta)}I_b^{s_b}
        \sim
        \cyc{k}{\alpha}\prodl_{a\in n(k,\alpha)}I_a^{s_{\theta(a)}}\cdot\cyc{k}{\beta}\prodl_{b\in
        n(k,\beta)}I_b^{s_{\inv{\theta}(b)}}
       \en
     \end{enumerate}
    \end{proposition}
   This theorem ensures \conv \clsfv of $\zt\wr\pern$ is totally determined
   by the structure of cycles with signature. We verify this
   statement by generalizing Young diagram technology. First,
   draw a {\it Young diagram with numbers and signatures} for each element $\prm\prodl_i{I_i^{s_i}}\in\zt\wr\pern$
   according to the decomposition Eq.(\ref{decompwnn}) by the
   following rules:
   \begin{enumerate}
    \item
    \label{r1}
     Plot Young diagram of the class in $\pern$ to which $\prm$ belongs and
     fill each row of this Young diagram with numbers in corresponding cycle by cyclic ordering from up-most box to down-most box.
    \item Draw a \slsv in the Young box if the number in this box is mapped to a minus-signed number.
   \end{enumerate}
   Secondly, partition elements in $\zt\wr\pern$ by their cycle structure in $\pern$ and Eq.(\ref{des}) guarantees elements belong to
   different partitions can not be \conv \id. Eq.(\ref{cyc}) implies that all the numbers that we filled by rule \ref{r1} are
   unnecessary, so smear them out and leave boxes and \sls es only. Within each row, Eq.(\ref{signat}) says that the positions of \sls es
   make no difference. What's more, in fact only that the total
   number of \sls es is even or odd distinguishes different
   classes. Therefore we regulate each row to contain zero or one
   \slsv at the bottom box. Eq.(\ref{exchg}) shows that we can
   not distinguish the case that one row without any \slsv (Mr. Zero) is put to the left
   to another row with one \slsv (Mr. One) from that Mr.Zero is to the right of Mr.One, if they have same
   cyclic length; thus we regulate that Mr.Zero shall always
   stand left to Mr.One. Therefore, conjugate classes of $\zt\wr\pern$ can be uniquely characterized by generalizing Young
   diagrams containing \sls es. Following Eq.(\ref{sh}), we represent \conv classes by
   \eq
   \label{syt}
    (\nu^+,\nu^-)=\prodl_{k=1}^n{(k^{\nu_{k}^{+}+\nu_{k}^{-}})}
   \en
   where $\nu_k^+$ is the number of Mr.Zero-type $k$-cycles and
   $\nu_k^-$ is that of Mr.One-type $k$-cycles, which satisfy $\nu_{k}^{+}+\nu_{k}^{-}=\nu_k$. It is not difficult to
   check some numerical properties of \conv classes of $\zt\wr\pern$.
   \begin{corollary}
    \begin{enumerate}
     \item Given a class $(\nu)$ in $\pern$, there are
      \eq
      \label{n1}
       \prodl_{k=1}^n {(1+\nu_k)}
      \en
      classes in $\zt\wr\pern$ which descend
      to $(\nu)$.
     \item The number of elements in a class $(\nu^+,\nu^-)$ is
      \eq
      \label{n2}
       N_{(\nu^+,\nu^-)}=N_{(\nu)}
       \prodl_{k=1}^n (C_{\nu_k}^{\nu_k^+} (\suml_{i=0}^{[{k\over 2}]}C_k^{2i})^{\nu_k^+} (\suml_{j=1}^{[{{k+1}\over 2}]} C_{k}^{2j-1})^{\nu_k^-})
      \en
      where $C_m^n$ is combinatorial number defined to be $m!/(n!(m-n)!)$.
     \item The order of a class $(\nu^+,\nu^-)$ is
      \eq
      \label{n3}
       lcm(\{k\cdot 2^{\dlt(\nu_k^-)}|\nu_k\neq 0\})
      \en
      where $\dlt(\nu_k^-)=0$, if $\nu_k^-=0$; $\dlt(\nu_k^-)=1$, if
      $\nu_k^->0$.
     \item Determinant (signature, parity) of a class
      \eq
      \label{n4}
       det((\nu^+,\nu^-))=\mos{\suml_{k=1}^n{\nu_k^-}}\cdot
       det((\nu))
      \en
      where $det((\nu))$ is the determinant of $(\nu)$ in $\pern$.
    \end{enumerate}
   \end{corollary}
   All in\idv \irr \rep s of $\ztnf$ can be expressed as
   \eq
   \label{zrp}
    \chi_{(s)}:=\otl_{p=1}^n{\chi_{\mns{s_p}}}
   \en
   in which $s_p\in \intg/2\intg, p=1,2,...,n$ and
   $\chi_{\mn},\chi_{(+)}$ are two \irrv \rep s of $\zt$ with
   $\chi_{(+)}$ being the unit \rep. Thus $\pi_{(s)}$ can be
   defined by Eq.(\ref{defpro}) and $\Pi(\ztnf)=\{\pi_{(s)}\}$.
   \footnote
   {$\pi_{(s)}$ satisfy $\pi_{(s)}\pi_{(s^\pr)}=\pi_{(s\cdot
   s^\pr)}$ where $(s\cdot s^\pr)(p)=s(p)s^\pr(p)$.}
   $\Pi(\ztnf)$ is divided into $n+1$ orbits under the
   $\pern$-action, namely
   $\Pi(\ztnf)=\coprod\limits_{p=0}^n{\Pi_p}$.
   For a given $p$, $\Pi_p$ consists of those $\pi_{(s)}$ who has $p$
   components in $(s)$ equal to $1$, other $n-p$ components equal
   to $0$; hence $|\Pi_p|=C_n^p$.
   Each $\orbt{p}{e}$ is specified to a $\pi_{(s)}$ with $s_p=0, p=1,2,...,n-p;
   s_p=1, p=n-p+1,...,n$, whose \stbsbv is just $\fixp$, denoted as
   $F_p$. Representatives of left-cosets in $S_n/F_p$ are written as $\prm_r$.  , then according
   to Eqs.(\ref{orbit})(\ref{main}) and Theorem \ref{rfm},
   \begin{proposition}(\Repv theory of $\zt\wr\pern$)
   \label{ron}
    \[
    \Prj{p}{\prm_r}{(\mu)i}{(\nu)j}=\prw{p}{\prm_r}{\mu}{i}{\nu}{j}
    \]
    give all in\idv \irrv \rep s of $\zt\wr\pern$ when $(p,(\mu),(\nu))$
    runs over its domain.
   \end{proposition}
   where $\orbt{p}{\prm_r}=\ad{\prm_r}{\orbt{p}{e}}$ whose $(s)$ will be denoted as
   $(s^{(p\prm_r)})$.
   \begin{corollary}
    \begin{enumerate}
     \item (Burside formula) $\suml_{(p,(\mu),(\nu))} (C_n^p \dm{\mu}\dm{\nu})^2=2^n n!$
     \item The number of \conv classes is
      $\suml_{(p,(\mu),(\nu))}1$.
     \item (\Repv matrix element)
      Given $\prm\prodl_q{I_q^{t_q}}\in \zt\wr\pern$,
      \[
       D_{(p,(\mu),(\nu))}(\prm\prodl_q{I_q^{t_q}})^{\prm_r^\pr i^\pr j^\pr}_{\prm_r
       ij}=\dlt^{\prm^\pr}_{\tld{\prm_r}(\prm\prm_r)} D_{(\mu)}(\prm_{(n-p)}(\prm\prm_r))^{i^\pr}_i
       D_{(\nu)}(\prm_p(\prm\prm_r))^{j^\pr}_j \prodl_q{\mns{s^{(p\prm_r)}_q t_q}}
      \]
     \item (Character)
      \[
       \chi_{(p,(\mu),(\nu))}(\prm\prodl_q{I_q^{t_q}})=\dlt^{\prm^\pr}_{\tld{\prm_r}(\prm\prm_r)}
       \chi_{(\mu)}(\prm_{(n-p)}(\prm\prm_r))
       \chi_{(\nu)}(\prm_p(\prm\prm_r))
       \prodl_q{\mns{s^{(p\prm_r)}_q t_q}}
      \]
    \end{enumerate}
   \end{corollary}
   where $\tld{\prm_r},\prm_{(n-p)},\prm_p$ map an element in
   $\pern$ to its decompositions according to $\pern/F_p$,
   $\per{n-p}$ and $\per{p}$ respectively.
   \subsection{Double group and spinor \rep}
   \label{doublegroup}
    Some fundamental facts of Clifford algebra are necessary for giving the definition and properties of
    double groups.
    Denote the Clifford algebra upon Euclidean space $V$ as
    $Cl(V)$; the isometry $x\mapsto -x$ on $V$ extends to an automorphism
    of $Cl(V)$ denoted by $x\mapsto
    \tilde{x}$ and referred to as the canonical
    automorphism of $Cl(V)$.
    We use $Cl^\ast(V)$ to denote the multiplicative group of
    invertible elements in $Cl(V)$ and the Pin group is the subgroup of $Cl^\ast(V)$ generated by
    unit vectors in $V$ , i.e.
    \[
     Pin(V):=\{a\in Cl^\ast(V): a=u_1\cdots u_r, u_j\in V,
     \|u_j\|=1\}
    \]
    Proofs of the following four statements can be found in
    \cite{harvey}.
    \begin{lemma}
    \label{reflection}
     If $u\in V$ is nonnull, then $R_u$, reflection along $u$, is
     given in terms of Clifford multiplication by
     \[
      R_u x=-uxu^{-1}, \forall x\in V
     \]
    \end{lemma}
    \begin{theorem}
     The sequence
     \[
      0\rightarrow \zt\rightarrow
      Pin(V)\stackrel{\wad}{\rightarrow} O(V)\rightarrow 1
     \]
     is exact, in which
     \[
      \wadf{a}{x}:=\tilde{a}xa^{-1}, \forall x\in Cl(V),
      a\in Pin(V)
     \]
    \end{theorem}
    We will usually write $\wad$ just by $\pi$ as a surjective homomorphism.
    \begin{proposition}
    \label{rep}
     $Cl(E^4)$, as an associative algebra with unit, is isomorphic
     to $M_2(\quaternion)$ where $\quaternion$ denotes quaternions.
    \end{proposition}
    \begin{lemma}
    \label{su4}
     Under the above algebra isomorphism, the image of $Pin(E^4)$
     is a subset of $SU(4)$.
    \end{lemma}
    Now we give the main definition of this paper.
    \begin{definition}
     Let $\ep$ be an injective homomorphism from a \gpv $G$ to $O(n)$, then the double group or the spin-extension of $G$
     with respect to $\ep$ is defined to be $D_n(G,\ep):=\pi^{-1}(G)$.
    \end{definition}
    An introduction to double groups in three dimension can be
    found in \cite{joshi}.
    Following elementary facts in the theory of group extension \cite{brown}, this diagram
    \[
     \CDalign{0&\CDto&\zt  &\CDto            &Pin(E^n)                        &\CDto^{\pi}&O(n)&\CDto&1 \cr
               &     &\CDeq&                 &\CDup&                &\CDup\CDrlabel{\epsilon}&        \cr
              0&\CDto&\zt  &\CDto            &\pi^{-1}(G)                     &\CDto           &G                       &\CDto&1  }
    \]
    is commutative. If $(G,\ep_1)\sim(G,\ep)$, there is
    \[
     \CDalign{0&\CDto&\zt  &\CDto^i         &\pi^{-1}((G_1,\epsilon_1))&\CDto^{\pi}       &(G,\epsilon_1)&\CDto&1 \cr
              &     &\CDeq&                &\CDdown                   &                  &\CDdown       &        \cr
             0&\CDto&\zt  &\CDto^{i^\prime}&\pi^{-1}((G_2,\epsilon_2))&\CDto^{\pi^\prime}&(G,\epsilon_2)&\CDto&1  }
    \]
    Note that the double group is not a universal object for a
    given abstract group $G$ but a special type of $\zt$-central extension of $G$
    subjected to the embedding
    $\ep$. For example, the results of doubling two $\zt$ subgroups in
    $O(2)$, $I:=\{1, \sigma\}, R:=\{1, R(\pi)\}$ where $\sigma$ denotes
    reflection along $y$-axes and $R(\pi)$ is the rotation over $\pi$, are $\pi^{-1}(I)\cong \zt\ot \zt$ while
    $\pi^{-1}(R)\cong Z_4$. Nevertheless, we will use symbol $\bar{G}$ to denote the double group
    at most cases where $n$ and $\epsilon$ are fixed. Meanwhile,
    symbol $\ebr$ is adopted to refer $-1$ in Clifford algebra and is called {\it{central element}}.\\
    \\
    Let $s: G\rightarrow \bar{G}, s.t. \pi s=Id_G$, namely $s$ is a cross-section of $\pi$.
    There is a property of the \conv classes of $\bar{G}$ which is
    easy to verify.
    \begin{lemma}
    \label{conjugat}
     Let $C$ be a \conv class in $G$, then either will $\pi^{-1}(C)$
     be one \conv class in $\bar{G}$ satisfying
     $\forall g\in C, s(g)\sim -s(g)$; or it will split into two \conv classes $C_1$, $C_2$ in
     $\bar{G}$ s.t. $\forall g\in C, s(g)\in C_1\Leftrightarrow -s(g)\in C_2$.
    \end{lemma}
    We will give a more deep result on the splitting of \conv
    classes when doubling $G$ to $\bar{G}$ in our another paper.\\
    \\
    Let $r$ be an \irrv \repv of $\bar{G}$ on $V$,
    then $r(-1)=\pm \unit$.
    \begin{definition}
     An \irrv \repv of $\bar{G}$ with $r(-1)=\unit$ is called a
     \svrepv of $G$ while an \irrv \repv with $r(-1)=-\unit$ is
     called a \spinrepv or \tvrepv of $G$.
    \end{definition}
    \begin{proposition}
    \label{repinduce}
     Let $\irrc{G}$ be the class of all in\idv \irrv
     \rep s of $G$ and $\irrc{G}^s$ be the class of all
     in\idv \svrep s of $G$,
     define $\phi:\irrc{G}\rightarrow \irrc{G}^s, r\mapsto r\circ
     \pi$. Then $\phi$ is a bijection.
    \end{proposition}
    \prf
     One can check: $r\circ \pi$ is a \repv of $\bar{G}$; if
     $r\cong r^\prime$, then $r\circ \pi\cong r^\prime\circ\pi$;
     that $r$ is \irrv implies that $r\circ\pi$ is \irrv and
     $r\circ\pi$ is single-valued. Therefore, $\phi$ is
     well-defined.
     If $r$ and $r^\prime$ are inequivalent, then $r\circ \pi$ and $r^\prime\circ\pi$ are
     two elements in $\irrc{G}^s$, namely $\phi$ is injective. To prove that $\phi$ is a
     surjection, consider any $\tilde{r}\in \irrc{G}^s:\bar{G}\rightarrow V$. Define
     $r:G\rightarrow V, g\mapsto\tilde{r}(s(g))$ where $s(g)$ is any element in $\pi^{-1}(g)$.
     One can check: $r$ is a well-defined map since $\tilde{r}$ is
     single-valued; $r$ is an \irrv \repv of $G$ on $V$,
     accordingly $r\in\irrc{G}$ and lastly, $\phi(r)=\tilde{r}$. So the result follows.\\
    \endprf\\
    This proposition says that all single-valued \rep s of $G$ which are part of inequivalent \irrv
    \rep s of $\bar{G}$ are completely determined by
    the \repv theory of $G$.
  \section{Structure of $\ohd$}
  \label{strandcc}
   \subsection{Structure of $\oh$}
   \label {sec.str}
    Following Proposition \ref{On},
    $\oh\cong \ztf\smdp S_4$; hence $|\oh|=384$. In
    point \gpv theory, rotation \subgv of $O_h$ is denoted as $O$
    and $\per{4}\cong \ztt\smdp S_3\cong O$ \cite{s}. We write the
    isomorphism explicitly. The structure of $\ztt \smdp S_3$ is given
    by four generators $\alpha, \beta, \eta, t$ and the relations
    \eqa
    \label{g1}
     \alpha^2=e, \beta^2=e, \alpha\beta=\beta\alpha\\
    \label{g2}
     t^3=e, \eta^2=e, \eta t=t^2\eta\\
    \label{g3}
     t\alpha=\alpha\beta t, t\beta=\alpha t,
     \eta \alpha=\beta \eta
    \ena
    and the isomorphisms are defined to be
    \eqann
     (12)(34)\leftrightarrow \alpha\leftrightarrow diag(-1,-1,1),
     (13)(24)\leftrightarrow \beta\leftrightarrow diag(1,-1,-1)\\
     (234)\leftrightarrow t\leftrightarrow
     \left(
      \begin{array}{lll}
       0&1&0\\
       0&0&1\\
       1&0&0
      \end{array}
     \right),
     (23)\leftrightarrow \eta\leftrightarrow
     \left(
      \begin{array}{lll}
       0&0&-1\\
       0&-1&0\\
       -1&0&0
      \end{array}
     \right)
    \enann
    The structure of $\ztf\smdp S_4$ is given by
    (\ref{g1})(\ref{g2})(\ref{g3}) together with (see Eq.(\ref{genr}))
    \eqa
    \label{o4g1}
     I_i^2=e, I_iI_j=I_jI_i, i,j=1..4, i\not= j\\
    \nn
     \alpha I_1=I_2\alpha, \alpha I_3=I_4\alpha\\
    \nn
     \beta I_1=I_3\beta,\beta I_2=I_4\beta\\
    \label{o4g2}
     t I_1=I_1 t,t I_2=I_4 t,t I_3=I_2 t,t I_4=I_3 t\\
    \nn
     \eta I_1=I_1\eta,\eta I_2=I_3\eta,\eta I_4=I_4\eta
    \ena
    The matrix form of $\oh$ is given by (see Eq.(\ref{cong}))
    \eqa
    \label{r41}
     (I_I)^j_k=\dlt^j_k(1-2\dlt^j_i), i,j,k=1,2,3,4\\
    \label{r42}
     \alpha\mapsto
      \left(
       \begin{array}{llll}
        0&1&0&0\\
        1&0&0&0\\
        0&0&0&1\\
        0&0&1&0
       \end{array}
      \right)
     \beta\mapsto
      \left(
       \begin{array}{llll}
        0&0&1&0\\
        0&0&0&1\\
        1&0&0&0\\
        0&1&0&0
       \end{array}
      \right)
     t\mapsto
      \left(
       \begin{array}{llll}
        1&0&0&0\\
        0&0&1&0\\
        0&0&0&1\\
        0&1&0&0
       \end{array}
      \right)
     \eta\mapsto
      \left(
       \begin{array}{llll}
        1&0&0&0\\
        0&0&1&0\\
        0&1&0&0\\
        0&0&0&1
       \end{array}
      \right)
    \ena
    In fact, if we introduce
    \[
     \gamma\mapsto
      \left(
       \begin{array}{llll}
        0&0&1&0\\
        0&1&0&0\\
        1&0&0&0\\
        0&0&0&1
       \end{array}
      \right)
    \]
    then the generating relations of $\oh$ can be reduced to a smaller
    set $\{I_i, \gamma, t|i=1,2,3,4\}$
    \footnote
    {
    In fact, we can demand a minimum generator set $\{ u,v\}$
    which we will not use here.
    \eqann
     u^2=e, v^6=e, \gamma=u, t=v^2\\
     (v^2u)^4=e, (v^3u)^4=e, u(v^2uv^3uv^4)=(v^2uv^3uv^4)u\\
     I_1=v^3, I_2=v^2uv^3uv^4, I_3=uv^3u, I_4=(v^4uv^3uv^2)
    \enann
    }
    whose generating relations are
    (\ref{o4g1})(\ref{o4g2}) together with
    \eqa
    \label{later1}
     \gamma^2=e, t^3=e,(t\gamma)^4=e\\
    \label{later2}
     \gamma I_1=I_3\gamma,\gamma I_2=I_2\gamma,\gamma I_4=I_4\gamma
    \ena
    while $\alpha=(t^2\gamma)^2,\beta=t\gamma t^2\gamma t, \eta=\gamma t\gamma
     t^2\gamma$.\\
     \\
    Applying the general results on \conv \clsfv of $\cgn$
    Eqs.(\ref{syt})(\ref{n1})...(\ref{n4}),
    we give the table of \conv classes of $\oh$ (see Tab. \ref{tab1}).
    \begin{table}[d]
    \caption{Conjugate Classes of $\ztn{4}\smdp\per{4}$}
    \label{tab1}
     \begin{tabbing}
      xxx\=xxxxxxxx\=xxxxxxxxxxxxxxxxx\=xxx\=xxxxx\=xxxxxxxxxx\=xxx\=xxxxxxxx\=xxxxxxxxxxxxxxxxx\=xxx\=xxxxx\=xxx \kill
      No\>SplitNo\>YoungDiagram\>ord\>num\>det \>No\>SplitNo\>YoungDiagram\>ord\>num\>det\\
      1\>1-1\>\ybxa{0}{0}{0}{0}\>1\>1\>1\>2\>1-2\>\ybxa{0}{0}{0}{10}\>2\>4\>-1\\
      3\>1-3\>\ybxa{0}{0}{10}{10}\>2\>6\>1\>4\>1-4\>\ybxa{0}{10}{10}{10}\>2\>4\>-1\\
      5\>1-5\>\ybxa{10}{10}{10}{10}\>2\>1\>1\\
      6\>2-1\>\ybxb{0}{0}{0}{0}\>2\>12\>-1\>7\>2-2\>\ybxb{0}{0}{0}{10}\>2\>24\>1\\
      8\>2-3\>\ybxb{10}{0}{0}{0}\>4\>12\>1\>9\>2-4\>\ybxb{0}{0}{10}{10}\>2\>12\>-1\\
      10\>2-5\>\ybxb{10}{0}{0}{10}\>4\>24\>-1\>11\>2-6\>\ybxb{10}{0}{10}{10}\>4\>12\>1\\
      12\>3-1\>\ybxc{0}{0}{0}{0}\>2\>12\>1\>13\>3-2\>\ybxc{0}{10}{0}{0}\>4\>24\>-1\\
      14\>3-3\>\ybxc{10}{10}{0}{0}\>4\>12\>1\\
      15\>4-1\>\ybxd{0}{0}{0}{0}\>3\>32\>1\>16\>4-2\>\ybxd{0}{0}{0}{10}\>6\>32\>-1\\
      17\>4-3\>\ybxd{10}{0}{0}{0}\>6\>32\>-1\>18\>4-4\>\ybxd{10}{0}{0}{10}\>6\>32\>1 \\
      19\>5-1\>\ybxe{0}{0}{0}{0}\>4\>48\>-1\>20\>5-2\>\ybxe{10}{0}{0}{0}\>8\>48\>1  \\
     \end{tabbing}
     "SplitNo" reflects the relation between the
     classes of $\zt\wr\per{4}$ and those of $\per{4}$. "ord" means order of each class.
     "num" is the number of elements in each class. "det" is the
     signature of each class. See
     Eqs.(\ref{n1})...(\ref{n4}).
    \end{table}
   \subsection{Construction of $\ohd$}
   \label{ohdstr}
    We will denote $s(g)$ still as $g$ for all $g\in G$.
    $\ohd$ is generated by the equations below.
    \begin{proposition}
    \label{gen}
     \eqa
     \label{o4bg1}
      I_i^2=-1, I_iI_j=-I_jI_i, i,j=1..4, i\not= j\\
     \label{o4bg2}
      \gamma^2=-1, t^3=-1, (t\gamma)^4=-1\\
     \label{o4bg3}
      {\underline{\gamma I_1=-I_3\gamma}}, \gamma I_2=-I_2\gamma, \gamma I_4=-I_4\gamma\\
     \label{o4bg4}
      tI_1=I_1t, {\underline{tI_2=I_4t}},{\underline{tI_3=I_2t}},
      {\underline{tI_4=I_3t}}
     \ena
    \end{proposition}
    \prf
     First
     Eqs.(\ref{o4bg1})...(\ref{o4bg4}) are
     valid. In fact, the standard orthogonal bases in $E^4$ satisfy Clifford
     relations
     $e_ie_j+e_je_i=-2\delta_{ij}$ which is
     equivalent to (\ref{o4bg1}); therefore, one can take
     $I_i=e_i$. Following lemma \ref{reflection}, we set
     $\gamma={1\over{\sqrt{2}}}(e_3-e_1)$ and check that
     (\ref{o4bg3}) is satisfied. Let $t={1\over 2}(1-e_2e_3+e_2e_4-e_3e_4)$ which is the product of
     ${1\over{\sqrt{2}}}(e_2-e_3)$ and
     ${1\over{\sqrt{2}}}(e_4-e_2)$, and (\ref{o4bg4}) can be
     verified. Finally, one can check that (\ref{o4bg2}) is also
     satisfied.\\
     \\
     Second, notice that above equations are just Eqs.(\ref{o4g1})(\ref{o4g2})(\ref{later1})(\ref{later2}),
     which generate $\oh$, twisted by a $\zt$ factor set.
     So due to the validity of the above equations, $\forall g\in \ohd
     $, either $g$ or $-g$ will be generated. But $-1$ can be
     generated. Therefore, the above equation set generations
     $\ohd$.\\
    \endprf
    We can give another proof of this result by proposition \ref{rep}.
    In fact, we introduce $\gamma$-matrices in $E^4$ as
    \[
     \gamma_i=\left(
      \begin{array}{cc}
       \zrt&i\sigma_i\\i\sigma_i&\zrt
      \end{array}\right),i=1,2,3;
     \gamma_4=\left(
      \begin{array}{cc}
       \zrt&-\unt\\ \unt&\zrt
      \end{array}\right)
    \]
    in which $\sigma_i$s stand for three Pauli matrices
    \[
     \sigma_1=\left(
      \begin{array}{cc}
       0&1\\1&0
      \end{array}\right),
     \sigma_2=\left(
      \begin{array}{cc}
       0&i\\-i&0
      \end{array}\right),
     \sigma_3=\left(
      \begin{array}{cc}
       1&0\\0&-1
      \end{array}\right)
    \]
    Note that our convention here has some difference with the usual
    one in physics.
    $\gamma_i (i=1..4)$ satisfy Clifford relations $\gamma_i\gamma_j
    +\gamma_j\gamma_i=-2\delta_{ij}\unf$ and $
     \gamma_i^\dag=-\gamma_i, \gamma_i \gamma_i^\dag=\unf,
     det(\gamma_i)=1$.\\
     \\
    We use $S(g)$ as the image of $s(g)$ in $M_2(\quaternion)$.
    Let
    \eq
    \label{L1}
     S(I_i)=\gamma_i,
     S(\gamma)={i\over\sqrt{2}}\cdot
      \left(
       \begin{array}{cccc}
        0&0&1&-1\\0&0&-1&-1\\1&-1&0&0\\-1&-1&0&0
       \end{array}
      \right),
     S(t)={e^{i\frac{7\pi}{4}}\over\sqrt{2}}\cdot
      \left(
       \begin{array}{cccc}
        1&-i&0&0\\1&i&0&0\\0&0&i&1\\0&0&-i&1
       \end{array}
      \right)
    \en
    then one can check that these matrices give correct images under $\widetilde{Ad}$ and satisfy
    the corresponding relations in (\ref{o4bg1})...(\ref{o4bg4}).
    It should be noticed that the $\widetilde{Ad}$-map condition can
    fix these matrices up to a non-vanishing scalar and that by
    using lemma \ref{su4}, the scalar can be fixed up to a $Z_4$
    uncertainty, namely if one find out a $S(g)$ then
    $iS(g),-S(g),-iS(g)$ will also work. Then one can figure out
    two of them by calculating the projections on the basis of
    $Cl(E^4)$ and ruling out those whose projections are pure
    imagine. \\
    \\
    We point out that the generating relations
    in proposition \ref{gen} are not unique, due to the canonical
    automorphism of $Cl(E^4)$. In fact from the second proof of
    this proposition, we have notified that at last there is still
    a $\zt$ uncertainty. Consequently, we can change the cross-section $s$ to
    another one $s^\prime$ by a "local" $\zt$ transformation and the
    underlined equations in
    Eqs.(\ref{o4bg1})...(\ref{o4bg4}) may gain or lose some
    $-1$-factors accordingly. Anyway, they are equivalent to the
    former ones.\\
    \\
    To classify the elements in $\ohd$, Lemma \ref{conjugat} will
    enable us to use the same symbols for the \conv classes of
    $\oh$ and to use a "$\pr$" for those splitting classes. Except for classes $1,8,14,15,20$
    which split into two classes for each, any other class in $\oh$ is lifted to one class. Therefore,
    there are totally  $25$ classes in $\ohd$ (see table \ref{tab2}).\\
    \begin{table}[hd]
    \caption{Conjugate Classes of $\ohd$}
    \label{tab2}
    \scriptsize
    \begin{tabular}{|l|r|r|r|r|r|r|r|r|r|r|r|r|r|r|r|r|r|r|r|r|r|r|r|r|r|}\hline
     No.&1&$1^\pr$&2&3&4&5&6&7&8&$8^\pr$&9&10&11&12&13&14&$14^\pr$&15&$15^\pr$&16&17&18&19&20&$20^\pr$\\ \hline
     num.&1&1&8&12&8&2&24&48&12&12&24&48&24&24&48&12&12&32&32&64&64&64&96&48&48\\ \hline
     ord.&1&2&4&4&2&2&4&4&8&8&2&8&8&4&8&4&4&6&3&12&6&6&4&8&8\\ \hline
    \end{tabular}
    \normalsize\\
    \\
    The labels of classes are descended from those of $\oh$ with
    "$\pr$" for those classes split when lifting into $\ohd$.
    \end{table}
  \section{\Rep s of $\ohd$}
  \label{repofohd}
   \subsection{Single-valued \rep s of $\oh$}
    Due to Theorem \ref{repinduce}, there are totally 20 in\idv \svrep s
    of $\oh$ corresponding to the 20 in\idv \irrv \rep s of $\oh$; the \repthv of $\oh$
    can be systematically solved by applying \ltgpv method
    (Theorem \ref{rfm}).\\
    \\
    All in\idv \irrv characters are listed in Table \ref{tab3}.
    \begin{table}[hd]
    \caption{Character Table of $\ztf$}
    \label{tab3}
    \scriptsize{
     \begin{tabular}{|l|c|c|c|c|c|c|c|c|c|c|c|c|c|c|c|c|}\hline
      $\ztfb$&$[e]$&$[I_1]$&$[I_2]$&$[I_3]$&$[I_4]$
      &$[I_{12}]$&$[I_{13}]$&$[I_{14}]$&$[I_{23}]$&$[I_{24}]$&$[I_{34}]$&$[I_{234}]$&$[I_{134}]$&$[I_{124}]$&$[I_{123}]$
      &$[I_{1234}]$\\ \hline
      $\chi_{0000}$&1&1&1&1&1&1&1&1&1&1&1&1&1&1&1&1\\ \hline
      $\chi_{0001}$&1&1&1&1&-1&1&1&-1&1&-1&-1&-1&-1&-1&1&-1\\ \hline
      $\chi_{0010}$&1&1&1&-1&1&1&-1&1&-1&1&-1&-1&-1&1&-1&-1\\ \hline
      $\chi_{0100}$&1&1&-1&1&1&-1&1&1&-1&-1&1&-1&1&-1&-1&-1\\ \hline
      $\chi_{1000}$&1&-1&1&1&1&-1&-1&-1&1&1&1&1&-1&-1&-1&-1\\ \hline
      $\chi_{0011}$&1&1&1&-1&-1&1&-1&-1&-1&-1&1&1&1&-1&-1&1\\ \hline
      $\chi_{0101}$&1&1&-1&1&-1&-1&1&-1&-1&1&-1&1&-1&1&-1&1\\ \hline
      $\chi_{1001}$&1&-1&1&1&-1&-1&-1&1&1&-1&-1&-1&1&1&-1&1\\ \hline
      $\chi_{0110}$&1&1&-1&-1&1&-1&-1&1&1&-1&-1&1&-1&-1&1&1\\ \hline
      $\chi_{1010}$&1&-1&1&-1&1&-1&1&-1&-1&1&-1&-1&1&-1&1&1\\ \hline
      $\chi_{1100}$&1&-1&-1&1&1&1&-1&-1&-1&-1&1&-1&-1&1&1&1\\ \hline
      $\chi_{1110}$&1&-1&-1&-1&1&1&1&-1&1&-1&-1&1&1&1&-1&-1\\ \hline
      $\chi_{1101}$&1&-1&-1&1&-1&1&-1&1&-1&1&-1&1&1&-1&1&-1\\ \hline
      $\chi_{1011}$&1&-1&1&-1&-1&-1&1&1&-1&-1&1&1&-1&1&1&-1\\ \hline
      $\chi_{0111}$&1&1&-1&-1&-1&-1&-1&-1&1&1&1&-1&1&1&1&-1\\ \hline
      $\chi_{1111}$&1&-1&-1&-1&-1&1&1&1&1&1&1&-1&-1&-1&-1&1\\ \hline
     \end{tabular}}
    \normalsize\\
    \\
    $I_{i_1i_2...I_a}:=I_{i_1}\cdot I_{i_2}\cdot ...\cdot I_{i_a}$. Irreducible characters are labeled as $\chi_{s_1s_2s_3s_4}, s_i\in
    \intg/2\intg$ (see Eq.(\ref{zrp})).
    \end{table}
    Following Theorem \ref{ron}, $\Pi(\ztf)$ are partitioned into orbits with index set defined in
    a physical convention $\indxst:=\{S,P,V,A,T\}$.
    \eqann
     \Pi_S=\{\pi_{0000}\}, F_S\cong S_4;\Pi_P=\{\pi_{1111}\}, F_P\cong S_4;\\
     \Pi_V=\{\pi_{0001},\pi_{0010},\pi_{0100},\pi_{1000}\}, F_V\cong S_3;\\
     \Pi_A=\{\pi_{1110},\pi_{1101},\pi_{1011},\pi_{0111}\}, F_A\cong S_3;\\
     \Pi_T=\{\pi_{0011},\pi_{0101},\pi_{1001},\pi_{0110},\pi_{1010},\pi_{1100}\}, F_T\cong \ztt
    \enann
    We will use $[\lam]$ instead of $(\nu)$ to denote Young
    diagrams where $[\lam]=[\lam_1\lam_2...\lam_n],
    \lam_k=\sum_{i=k}^n{\nu_i}$.\\
    \\
    \orb S\\
    \label{orb1}
     All in\idv \irrv \rep s of $S_4$ is labeled by
     $[4],[31],[2^2],[21^2],[1^4]$; accordingly,
     \[
      \Pi_S\cdot([4],[31],[2^2],[21^2],[1^4])
     \]
     provide two one-\diml, one two-\dimlv and two three-\dimlv
     \rep s. As for \repv matrices, all $I_i,i=1..4$ are mapped to identity, while $\alpha,
     \beta,t,\eta$ take the same matrix form as they have in $S_4$, i.e.
     $\Pi_S\cdot[4]:
       I_i,\alpha,\beta,t,\eta\rightarrow 1$;
     $\Pi_S\cdot[1^4]:
       I_i,\alpha,\beta,t\rightarrow 1,\eta\rightarrow -1$;
     \[
      \Pi_S\cdot[2^2]:
       I_i,\alpha,\beta\rightarrow\unt,
       t\rightarrow
        \left(
         \begin{array}{cc}
          \omega&0\\0&\omega^2
         \end{array}
        \right),
       \eta\rightarrow
        \left(
         \begin{array}{cc}
          0&1\\1&0
         \end{array}
        \right);
      \]
      \[
      \Pi_S\cdot[31]:
       I_i\rightarrow\unth,
       \alpha\rightarrow
        \left(
         \begin{array}{ccc}
          -1&0&0\\0&1&0\\0&0&-1
         \end{array}
        \right),
       \beta\rightarrow
        \left(
         \begin{array}{ccc}
          -1&0&0\\0&-1&0\\0&0&1
         \end{array}
        \right),
       t\rightarrow
        \left(
         \begin{array}{ccc}
          0&0&1\\1&0&0\\0&1&0
         \end{array}
        \right),
       \eta\rightarrow
        \left(
         \begin{array}{ccc}
          1&0&0\\0&0&1\\0&1&0
         \end{array}
        \right);
      \]
    $\Pi_S\cdot[21^2]$: $I_i, \alpha, \beta, t$ take the same form
    of $\Pi_S\cdot[31]$ and $\eta$ gains a minus sign compared to
    $\Pi_S\cdot[31]$.\\
    \\
  \orb P\\
   \[
    \Pi_P\cdot([4],[31],[2^2],[21^2],[1^4])
   \]
   The only difference from \orb S is that $I_i$ are mapped to
   $-\unit$.\\
   \\
  \orb V\\
  \label{orb3}
   All in\idv \irrv \rep s of $S_3$ can be written as $[3],[21],[1^3]$
   and it has no difficulty,
   using our generating relations, to check
   \[
    \alpha\pi_{1000}\alpha^{-1}=\pi_{0100},
    \eta\pi_{0100}\eta^{-1}=\pi_{0010},
    \alpha\pi_{0010}\alpha^{-1}=\pi_{0001}
   \]
   Hence, this orbit gives two four-\dimlv \rep s and one eight-\dimlv
   \rep.
   \[
    \Pi_V\cdot ([3],[21],[1^3])=(e,\alpha,\beta\eta,\alpha\beta\eta)\cdot\pi_{1000}\cdot ([3],[21],[1^3])
   \]
   The \repv matrices of $\Pi_V\cdot [3]$ are coincident with those in Eqs.(\ref{r41})(\ref{r42}).
   \Repv matrices of $\Pi_V\cdot [1^3]$ are the same as those in $\Pi_V\cdot [3]$,
   except that $\eta$ picking on a minus sign. \\
   \\
   $\Pi_V\cdot [21]:$
    \[
     e_i\rightarrow
      \left(
       \begin{array}{cc}
        \Pi_V\cdot [3](e_i)&\zrf\\
        \zrf&\Pi_V\cdot [3](e_i)
       \end{array}
      \right),
    \]
    \eqann
     \alpha\rightarrow
      \left(
       \begin{array}{cc}
        \Pi_V\cdot [3](\alpha)&\zrf\\
        \zrf&\Pi_V\cdot [3](\alpha)
       \end{array}
      \right),
     \beta\rightarrow
      \left(
       \begin{array}{cc}
        \zrf&\Pi_V\cdot [3](\beta)\\
        \Pi_V\cdot [3](\beta)&\zrf
       \end{array}
      \right),\\
     t\rightarrow
      \left(
       \begin{array}{cc}
        \omega\cdot\Pi_V\cdot [3](t)&\zrf\\
        \zrf&\omega^2\cdot\Pi_V\cdot [3](t)
        \end{array}
      \right),
     \eta\rightarrow
      \left(
       \begin{array}{cc}
        \zrf&\Pi_V\cdot [3](\eta)\\
        \Pi_V\cdot [3](\eta)&\zrf
       \end{array}
      \right)
    \enann
  \orb A
  \label{orb4}\\
   Similar to \orb V, there are two four-\dimlv \rep s and one eight-\dimlv
   \rep.
   \[
    \Pi_A\cdot ([3],[21],[1^3])=(e,\alpha,\beta\eta,\alpha\beta\eta)\cdot\pi_{0111}\cdot([3],[21],[1^3])
   \]
   while the \repv matrices for $I_i$ pick on a minus sign, without changing the
   others.\\
   \\
  \orb T\\
   The \stbsbv $F_T$ leaving $\pi_{0110}$ invariant is
   $\{e,\eta,\alpha\beta,\alpha\beta\eta\}$ with
   four one-dimensional \irrv \rep s, denoted by
   $\pi_{(a,b)},a,b=0,1$. Therefore, there are four six-\dimlv \rep s given by
   this orbit. Notice that
   \eqann
    \alpha\pi_{0110}\alpha^{-1}=\pi_{1001},
    t\pi_{1001}t^{-1}=\pi_{1010},
    t\pi_{1010}t^{-1}=\pi_{1100},\\
    \alpha\pi_{1010}\alpha^{-1}=\pi_{0101},
    \beta\pi_{1100}\beta^{-1}=\pi_{0011},
   \enann
   the four \rep s can be labeled as
   \[
    \Pi_T\cdot(\pi_{00},\pi_{01},\pi_{10},\pi_{11})=(e,\alpha,\alpha\beta t,\beta t^2,\beta
    t,t^2)\cdot \pi_{0110}
    \cdot(\pi_{00}+\pi_{01}+\pi_{10}+\pi_{11})
   \]
   Then we enumerate the matrices for the four \rep s.
   \[
    \Pi_T \cdot\pi_{00}:
     I_1\rightarrow diag(1,-1,-1,-1,1,1),I_2\rightarrow diag(-1,1,1,-1,-1,1),
   \]
   \[
    I_3\rightarrow diag(-1,1,-1,1,1,-1),I_4\rightarrow diag(1,-1,1,1,-1,-1),
   \]
   \[
    \alpha\rightarrow
      \left(
       \begin{array}{cccccc}
        0&1&0&0&0&0\\
        1&0&0&0&0&0\\
        0&0&0&0&1&0\\
        0&0&0&1&0&0\\
        0&0&1&0&0&0\\
        0&0&0&0&0&1
       \end{array}
      \right)
     \beta\rightarrow
      \left(
       \begin{array}{cccccc}
        0&1&0&0&0&0\\
        1&0&0&0&0&0\\
        0&0&1&0&0&0\\
        0&0&0&0&0&1\\
        0&0&0&0&1&0\\
        0&0&0&1&0&0
       \end{array}
      \right)
    \]
    \[
     t\rightarrow
      \left(
       \begin{array}{cccccc}
        0&0&0&0&0&1\\
        0&0&0&1&0&0\\
        0&1&0&0&0&0\\
        0&0&1&0&0&0\\
        1&0&0&0&0&0\\
        0&0&0&0&1&0
       \end{array}
      \right)
     \eta\rightarrow
      \left(
       \begin{array}{cccccc}
        1&0&0&0&0&0\\
        0&1&0&0&0&0\\
        0&0&0&1&0&0\\
        0&0&1&0&0&0\\
        0&0&0&0&0&1\\
        0&0&0&0&1&0
       \end{array}
      \right)
   \]
   \[
    \Pi_T \cdot\pi_{01}:
     I_1\rightarrow \Pi_2 \cdot\pi_{00}(I_1),
     I_2\rightarrow \Pi_T \cdot\pi_{00}(I_2),
     I_3\rightarrow \Pi_T \cdot\pi_{00}(I_3),
     I_4\rightarrow \Pi_T \cdot\pi_{00}(I_4)
   \]
   \[
     \alpha\rightarrow
      \left(
       \begin{array}{cccccc}
        0&1&0&0&0&0\\
        1&0&0&0&0&0\\
        0&0&0&0&1&0\\
        0&0&0&-1&0&0\\
        0&0&1&0&0&0\\
        0&0&0&0&0&-1
       \end{array}
      \right)
     \beta\rightarrow
      \left(
       \begin{array}{cccccc}
        0&-1&0&0&0&0\\
        -1&0&0&0&0&0\\
        0&0&-1&0&0&0\\
        0&0&0&0&0&1\\
        0&0&0&0&-1&0\\
        0&0&0&1&0&0
       \end{array}
      \right)
   \]
   \[
     t\rightarrow
      \left(
       \begin{array}{cccccc}
        0&0&0&0&0&1\\
        0&0&0&1&0&0\\
        0&1&0&0&0&0\\
        0&0&1&0&0&0\\
        -1&0&0&0&0&0\\
        0&0&0&0&-1&0
       \end{array}
      \right)
     \eta\rightarrow
      \left(
       \begin{array}{cccccc}
        1&0&0&0&0&0\\
        0&-1&0&0&0&0\\
        0&0&0&-1&0&0\\
        0&0&-1&0&0&0\\
        0&0&0&0&0&-1\\
        0&0&0&0&-1&0
       \end{array}
      \right)
    \]
    \[
    \Pi_T \cdot\pi_{10}:
     I_1\rightarrow \Pi_T \cdot\pi_{00}(I_1),
     I_2\rightarrow \Pi_T \cdot\pi_{00}(I_2),
     I_3\rightarrow \Pi_T \cdot\pi_{00}(I_3),
     I_4\rightarrow \Pi_T \cdot\pi_{00}(I_4)
    \]
    \[
     \alpha\rightarrow \Pi_T \cdot\pi_{00}(\alpha),
     \beta\rightarrow \Pi_T \cdot\pi_{00}(\beta),
     t\rightarrow \Pi_T \cdot\pi_{00}(t),
     \eta\rightarrow (-1)\cdot\Pi_T \cdot\pi_{00}(\eta)
    \]
    \[
    \Pi_T \cdot\pi_{11}:
     I_1\rightarrow \Pi_T \cdot\pi_{01}(I_1),
     I_2\rightarrow \Pi_T \cdot\pi_{01}(I_2),
     I_3\rightarrow \Pi_T \cdot\pi_{01}(I_3),
     I_4\rightarrow \Pi_T \cdot\pi_{01}(I_4)
    \]
    \[
     \alpha\rightarrow \Pi_T \cdot\pi_{01}(\alpha),
     \beta\rightarrow \Pi_T \cdot\pi_{01}(\beta),
     t\rightarrow \Pi_T \cdot\pi_{01}(t),
     \eta\rightarrow (-1)\cdot\Pi_T \cdot\pi_{01}(\eta)
    \]
  Here we find all 20 inequivalent \irrv \rep s corresponding
  to the 20 \conv classes of $\oh$, which satisfy Burside formula
  \[
   2\times(1^2+1^2+2^2+3^2+3^2)+2\times(4^2+4^2+8^2)+4\times 6^2
   =384
  \]
  Following proposition \ref{repinduce}, we have found all of the
  single-valued \rep s of $\oh$.
 \subsection{Spinor \rep s of $\oh$}
  Notice the following facts that $\ov{\ztf}\lhd\ohd$, $\ohd/\ov{\ztf}\cong
  \per{4}$ and Eqs.(\ref{L1}) generate a \spinrepv of
  $\oh$ which is denoted still as $S$; what's more, its restriction to $\ov{\ztf}$
  is also a \tvrepv of $\ztf$. These facts ensure two conditions in Theorem
  \ref{clifford}. To apply Theorem \ref{clifford} to deduce \spinrep s of $\oh$, we develop a
  calculation method. The matrices of a \spinrepv of $\oh$ for $I_i,\gamma,t$, denoted
  as $\tld{S}(I_i),\tld{S}(\gamma),\tld{S}(t)$, can be decomposed as
  \eqann
   \tld{S}(I_i)=S(I_i)\ot\unit,i=1,3,4;\tld{S}(I_2)=-S(I_2)\ot\unit;\\
   \tld{S}(\gamma)=\Gamma\ot \tld{\gamma};
   \tld{S}(t)=T\ot\tilde{t}
  \enann
  where $\Gamma,T$ and $S(I_i)$ act on the same module,
  $\tilde{\gamma},\tilde{t}$ have \dfn{the same
  texture (zero matrix elements) of the \repv matrices of five in\idv \irrv \rep s of $S_4$} (
  the minus added before $S(I_2)$ is for a physical convention).
  There are five \spinrep s of dimension 4,4,8,12 and 12 respectively and
  the second half of Burside formula is satisfied.
  \[
   4^2+4^2+8^2+12^2+12^2=384
  \]
  Corresponding the generating equations
  (\ref{o4bg1})...(\ref{o4bg4}), there are
  a system of matrix equations.
  \eq
  \label{4deq1}
   \tld{S}(\gamma)^2=\tld{S}(t)^3=-\unit,(\tld{S}(\gamma)\tld{S}(t))^4=-\unit
  \en
  \eq
  \label{4deq2}
   \tld{S}(\gamma)\tld{S}(I_2)=-\tld{S}(I_2)\tld{S}(\gamma),
   \tld{S}(\gamma)\tld{S}(I_4)=-\tld{S}(I_4)\tld{S}(\gamma),
   \tld{S}(\gamma)\tld{S}(I_1)=-\tld{S}(I_3)\tld{S}(\gamma),
  \en
  \eq
  \label{4deq3}
   \tld{S}(t)\tld{S}(I_1)=\tld{S}(I_1)\tld{S}(t),
   \tld{S}(t)\tld{S}(I_2)=-\tld{S}(I_4)\tld{S}(t),
   \tld{S}(t)\tld{S}(I_3)=-\tld{S}(I_2)\tld{S}(t),
   \tld{S}(t)\tld{S}(I_4)=\tld{S}(I_3)\tld{S}(t),
  \en
  plus a unitary condition
  \eq
  \label{4deq4}
   \tld{S}(\gamma)^{\dagger}\tld{S}(\gamma)=\unit,
   \tld{S}(t)^{\dagger}\tld{S}(t)=\unit
  \en
  Note that we add a minus sign to the second and the third equations in
  Eq.(\ref{4deq3}) compared with Eq.(\ref{o4bg4}) according to the same physics convention, though they
  are completely equivalent.\\
  \\
  Solving Eqs.(\ref{4deq1})...(\ref{4deq4}) for four-\dimlv case gives two solutions
  \[
   \b{4}_{+}:
    \tld{S}(\gamma)=\Gamma\cdot \tilde{\gamma}_{+},
    \Gamma={1\over\sqrt{2}}\cdot
     \left(
      \begin{array}{cccc}
       0&0&1&-1\\0&0&-1&-1\\1&-1&0&0\\-1&-1&0&0
      \end{array}
     \right),
    \tilde{\gamma}_{+}=i
  \]
  \[
    \tld{S}(t)=T\cdot\tilde{t},
    T={1\over\sqrt{2}}\cdot
     \left(
      \begin{array}{cccc}
       1&-i&0&0\\1&i&0&0\\0&0&i&1\\0&0&-i&1
      \end{array}
     \right),
    \tilde{t}=e^{i\frac{7\pi}{4}}
  \]
  \[
   \b{4}_{-}:
    \tld{S}(\gamma)=\Gamma\cdot \tilde{\gamma}_{-},
    \tilde{\gamma}_{-}=-i,
    \tld{S}(t)=T\cdot\tilde{t}
  \]
  Note that $\b{4}_{+}$ is just the \repv $S$ with $\tld{S}(I_2)=-S(I_2)$.\\
  \\
  As for eight-\dimlv case we can suppose
  \[
   \tld{\gamma}=
    \left(
     \begin{array}{cc}
      0&\tilde{c}\\\tilde{d}&0
     \end{array}
    \right),
   \tld{t}=
    \left(
     \begin{array}{cc}
      \tilde{a}&0\\0&\tilde{d}
     \end{array}
    \right)
  \]
   The solution $\b{8}$ is given by
  \[
   \tld{c}=e^{i\frac{\pi}{3}},\tld{d}=e^{i\frac{2\pi}{3}},
   \tld{a}=e^{i\frac{5\pi}{12}},\tld{d}=e^{i\frac{13\pi}{12}}
  \]
  Finally, we set for the twelve-\dimlv case
  \[
   \tld{\gamma}=
    \left(
     \begin{array}{ccc}
      0&0&\tld{z}\\0&\tld{y}&0\\ \tld{x}&0&0
     \end{array}
    \right),
   \tld{t}=
    \left(
     \begin{array}{ccc}
      0&0&\tld{n}\\ \tld{l}&0&0\\ 0&\tld{m}&0
     \end{array}
    \right)
  \]
  Such that
  \eqann
   \b{12}_+:
    \tld{x}=1,\tld{y}=i,\tld{z}=-1,
    \tld{l}=1,\tld{m}=1,\tld{n}=e^{i\frac{5\pi}{4}}\\
   \b{12}_-:
    \tld{x}=1,\tld{y}=-i,\tld{z}=-1,
    \tld{l}=1,\tld{m}=-1,\tld{n}=e^{i\frac{\pi}{4}}
  \enann
  So far, we obtain all in\idv \irrv \rep s of $\ohd$ and we summarize our results in Table \ref{tabch}.
  \small
  \begin{table}[hd]
  \caption{Character Table of $\ohd$}
  \label{tabch}
  \tabcolsep .7pt
  \begin{tabular}{|l||c|c|c|c|c|c|c|c|c|c|c|c|c|c|c|c|c|c|c|c||c|c|c|c|c|}\hline
   &\multicolumn{5}{c|}{$\Pi_S$}&\multicolumn{5}{c|}{$\Pi_P$}&
   \multicolumn{3}{c|}{$\Pi_V$}&\multicolumn{3}{c|}{$\Pi_A$}&\multicolumn{4}{c|}{$\Pi_T$}&\multicolumn{5}{c|}{spinor
   rep.}\\ \cline{2-26}
   &$[4]$&$[1^4]$&$[2^2]$&$[31]$&$[21^2]$&$[4]$&$[1^4]$&$[2^2]$&$[31]$&$[21^2]$
   &$[3]$&$[1^3]$&$[21]$&$[3]$&$[1^3]$&$[21]$&$\pi_{00}$&$\pi_{01}$&$\pi_{10}$&$\pi_{11}$
      &$\b{4}_+$&$\b{4}_-$&$\b{8}$&$\b{12}_+$&$\b{12}_-$\\ \hline\hline
                 1&1&1&2&3&3&1&1&2&3&3&4&4&8&4&4&8&6&6&6&6&4&4&8&12&12\\ \hline
                 $1^\pr$&1&1&2&3&3&1&1&2&3&3&4&4&8&4&4&8&6&6&6&6&-4&-4&-8&-12&-12\\ \hline
                 2&1&1&2&3&3&-1&-1&-2&-3&-3&2&2&4&-2&-2&-4&0&0&0&0&0&0&0&0&0\\ \hline
                 3&1&1&2&3&3&1&1&2&3&3&0&0&0&0&0&0&-2&-2&-2&-2&0&0&0&0&0\\ \hline
                 4&1&1&2&3&3&-1&-1&-2&-3&-3&-2&-2&-4&2&2&4&0&0&0&0&0&0&0&0&0\\ \hline
                 5&1&1&2&3&3&1&1&2&3&3&-4&-4&-8&-4&-4&-8&6&6&6&6&0&0&0&0&0\\ \hline
                 6&1&-1&0&1&-1&1&-1&0&1&-1&2&-2&0&2&-2&0&2&0&-2&0&0&0&0&0&0\\ \hline
                 7&1&-1&0&1&-1&-1&1&0&-1&1&0&0&0&0&0&0&0&2&0&-2&0&0&0&0&0\\ \hline
                 8&1&-1&0&1&-1&-1&1&0&-1&1&2&-2&0&-2&2&0&0&-2&0&2&$-2\sqrt{2}$&$2\sqrt{2}$&0&$-2\sqrt{2}$&$2\sqrt{2}$\\ \hline
         $8^\pr$&1&-1&0&1&-1&-1&1&0&-1&1&2&-2&0&-2&2&0&0&-2&0&2&$2\sqrt{2}$&$-2\sqrt{2}$&0&$2\sqrt{2}$&$-2\sqrt{2}$\\ \hline
                 9&1&-1&0&1&-1&1&-1&0&1&-1&-2&2&0&-2&2&0&2&0&-2&0&0&0&0&0&0\\ \hline
                10&1&-1&0&1&-1&1&-1&0&1&-1&0&0&0&0&0&0&-2&0&2&0&0&0&0&0&0\\ \hline
                11&1&-1&0&1&-1&-1&1&0&-1&1&-2&2&0&2&-2&0&0&-2&0&2&0&0&0&0&0\\ \hline
                12&1&1&2&-1&-1&1&1&2&-1&-1&0&0&0&0&0&0&2&-2&2&-2&0&0&0&0&0\\ \hline
                13&1&1&2&-1&-1&-1&-1&-2&1&1&0&0&0&0&0&0&0&0&0&0&0&0&0&0&0\\ \hline
                14&1&1&2&-1&-1&1&1&2&-1&-1&0&0&0&0&0&0&-2&2&-2&2&2&2&4&-2&-2\\ \hline
   $14^\pr$&1&1&2&-1&-1&1&1&2&-1&-1&0&0&0&0&0&0&-2&2&-2&2&-2&-2&-4&2&2\\ \hline
                15&1&1&-1&0&0&1&1&-1&0&0&1&1&-1&1&1&-1&0&0&0&0&2&2&-2&0&0\\ \hline
   $15^\pr$&1&1&-1&0&0&1&1&-1&0&0&1&1&-1&1&1&-1&0&0&0&0&-2&-2&2&0&0\\ \hline
                16&1&1&-1&0&0&-1&-1&1&0&0&-1&-1&1&1&1&-1&0&0&0&0&0&0&0&0&0\\ \hline
                17&1&1&-1&0&0&-1&-1&1&0&0&1&1&-1&-1&-1&1&0&0&0&0&0&0&0&0&0\\ \hline
                18&1&1&-1&0&0&1&1&-1&0&0&-1&-1&1&-1&-1&1&0&0&0&0&0&0&0&0&0\\ \hline
                19&1&-1&0&-1&1&1&-1&0&-1&1&0&0&0&0&0&0&0&0&0&0&0&0&0&0&0\\ \hline
                20&1&-1&0&-1&1&-1&1&0&1&-1&0&0&0&0&0&0&0&0&0&0&$\sqrt{2}$&$-\sqrt{2}$&0&$-\sqrt{2}$&$\sqrt{2}$\\ \hline
   $20^\pr$&1&-1&0&-1&1&-1&1&0&1&-1&0&0&0&0&0&0&0&0&0&0&$-\sqrt{2}$&$\sqrt{2}$&0&$\sqrt{2}$&$-\sqrt{2}$\\ \hline
  \end{tabular}
  \end{table}
  \normalsize\\
  \\
   {\bf Acknowledgements}\\
    This work was supported by Climb-Up (Pan Deng) Project of
    Department of Science and Technology in China, Chinese
    National Science Foundation and Doctoral Programme Foundation
    of Institution of Higher Education in China.
    One of the authors J.D. is grateful to Dr. L-G. Jin for his advice on this paper.
  \appendix
  \setcounter{equation}{0}
  \renewcommand{\theequation}{A-\arabic{equation}}
  \setcounter{definition}{0}
  \renewcommand{\thedefinition}{A-\arabic{definition}}
  \setcounter{lemma}{0}
  \renewcommand{\thelemma}{A-\arabic{lemma}}
  \setcounter{proposition}{0}
  \renewcommand{\theproposition}{A-\arabic{proposition}}
  \setcounter{corollary}{0}
  \renewcommand{\thecorollary}{A-\arabic{corollary}}
  \section{Fundamental lemma of n-dimensional Euclidean geometry}
  \label{app2}
   \begin{lemma}
   \label{fundament}
   (Weak form)
    Let $p_i, i=0,1,2,...,n$ be $n$ points in
    n-dimensional Euclidean space $\eun$ which are non-collinear
    and given $n$ non-negative real numbers $d_i, i=0,1,2,...,n$, then there exists
    at most one point $p\in \eun$ s.t. $d(p,p_i)=d_i$.
   \end{lemma}
   \prf
    Without losing generality, set $p_0=(0,0,...,0)$ and
    understand $p, p_i, i=1,2,...,n$ as vectors in $\eun$.
    Consider equation set
    \eqa
    \label{disf}
     (p-p_i, p-p_i)&=&d_i^2, i=1,2,...,n\\
    \label{disf1}
     (p,p)&=&d_0^2
    \ena
    Substitute (\ref{disf1}) into (\ref{disf})
    \eq
    \label{disf2}
     (p_i,p)=\hf(d_0^2-d_i^2+(p_i,p_i)), i=1,2,...,n
    \en
    The non-collinearity implies that (\ref{disf2}) has a solution
    $p$. The weak form of {\it fundamental lemma of Euclidean
    geometry} follows. \\
   \endprf
  

\begin{thebibliography}{99}
   \bibitem{bc} J. L. Birman and Li-Ching Chen, "Irreducible vector and ray \rep s
    of some cubic crystal point \gp s in four \dim", J. Math. Phys. {\bf
    12}(1971)2454
   \bibitem{young} A. Young, "On quantitative substitutional analysis (Fifth paper)", Proc. London
    Math. Soc. (2){\bf 31}(1930)273-288
   \bibitem{sp1} W. Specht, "Eine Verallgemeinerung der symmetrischen Gruppe", Schriften Berlin {\bf 1}(1932)1-32
   \bibitem{sp2} W. Specht, "Eine Verallgemeinerung der Permutationsgruppen", Math. Z. {\bf 37}(1933)321-341
   \bibitem{kerber} Adalbert Kerber, {\it{Representations of Permutation Groups
    I}}, Spinger-Verlag (1971)
   \bibitem{wilson} K. G. Wilson, Phys. Rev. D {\bf 14}(1974)2445
   \bibitem{baake} M. Baake, B. Gem\"{u}nden and R. Oedingen, "Structure and \rep s of the
    symmetry group of the four-\dimv cube", J. Math. Phys. {\bf
    23}(1982)944-953
   \bibitem{mandula} J. E. Mandula, G. Zweig and J. Govaerts, "\Rep s of the rotation reflection
   symmetry group of the four-\dimlv cubic lattice", Nucl. Phys. B {\bf
   228}(1983)91-108
   \bibitem{cr} Charles W.Curtis and Irving Reiner, {\it{Methods of Representation
    Theory}} vol.I, John Wiley \& Sons Inc. (1981)
   \bibitem{s} Jean-Pierre Serre, {\it{Linear Representations of Finite
    Groups}}, Springer-Verlag (1977)
   \bibitem{cs} Xi-Hua Cao, Qian-Yi Shi, {\it{\Repv Theory of Finite
    Groups}}, Higher Education Press, Beijing (1992)
   \bibitem{qiu} Wei-Sheng Qiu, {\it{\Repv Theory of Finite Groups and Compact
    Groups}}, Peking Universiry Press, Beijing (1997)
   \bibitem{clif} A. H. Clifford, "\Rep s induced in an invariant \subg", Ann. of Math. (2){\bf
    38}(1937)533-550
   \bibitem{hs} Qi-Zhi Han, Hong-Zhou Sun, {\it{Group Theory}}, Peking University Press, Beijing (1987)
   \bibitem{harvey} F. Reese Harvey, {\it{Spinors and
    Calibrations}}, academic press (1990)
   \bibitem{joshi} A. W. Joshi, {\it Elements of Group Theory for Physicists},
    John Wiley (1977)
   \bibitem{brown} Kenneth S. Brown, {\it{Cohomology of Groups}},
    springer-verlag (1982)
  \end{thebibliography}
 \end{document}